\documentclass[draft]{llncs}
 
\usepackage{amsmath}
\usepackage{amssymb}
\usepackage{stmaryrd}
\usepackage{url}
\usepackage{ifthen}

\newboolean{short}
\setboolean{short}{false}

\spnewtheorem*{delayedproof}{Proof}{\bfseries}{\rmfamily}

\begin{document}
\ifthenelse{\boolean{short}}
{%
\title{Widening Operators for Weakly-Relational Numeric Abstractions
       (Extended Abstract)%
\thanks{This work has been partly supported by projects
  ``Constraint Based Verification of Reactive Systems''
  and
  ``AIDA --- Abstract Interpretation Design and Applications''
  and
  by the Royal Society International Joint Project --- 2004/R1 Europe (ESEP).
}
}
}{%
\title{Widening Operators \\ for Weakly-Relational Numeric Abstractions%
\thanks{This work has been partly supported by projects
  ``Constraint Based Verification of Reactive Systems''
  and
  ``AIDA --- Abstract Interpretation Design and Applications''
  and
  by the Royal Society International Joint Project --- 2004/R1 Europe (ESEP).
}
}
\subtitle{(Extended Abstract)}
}
\author{Roberto Bagnara\inst{1}
\and Patricia M. Hill\inst{2}
\and Elena Mazzi\inst{1}
\and Enea Zaffanella\inst{1}
}
\authorrunning{R.~Bagnara,
               P.~M.~Hill,
               E.~Mazzi,
               E.~Zaffanella,
}
\tocauthor{Roberto Bagnara (University of Parma),
           Patricia M. Hill (University of Leeds),
           Elena Mazzi (University of Parma),
           Enea Zaffanella (University of Parma)
}

\institute{
    Department of Mathematics,
    University of Parma,
    Italy \\
    \email{\{bagnara,%
             mazzi,%
             zaffanella\}@cs.unipr.it}
\and
    School of Computing,
    University of Leeds,
    UK \\
    \email{hill@comp.leeds.ac.uk}
}
                                                                                
\maketitle

\ifthenelse{\boolean{short}}
{%
\enlargethispage{4mm}
}{%
\section{Introduction}
}
In recent years there has been a lot of interest in the definition of
so-called \emph{weakly-relational} numeric domains, whose complexity
and precision are in between the (non-relational) abstract domain of
intervals~\cite{CousotC77} and the (relational) abstract domain of
convex polyhedra~\cite{CousotH78}.
The first weakly-relational domain proposed in the literature is based
on systems of constraints of the form $x - y \leq c$ and $\pm x \leq c$,
typically represented by Difference-Bound Matrices (DBMs).  Even
though DBMs have a long tradition in Computer Science, their use in
the Abstract Interpretation field is quite recent. The idea of
defining an abstract domain of DBMs was put forward
in~\cite{Bagnara97th}, where these constraints were called
\emph{bounded differences}.  An independent application can be found
in~\cite{ShahamKS00}, where an abstract domain of transitively closed
DBMs is defined.  In this case, the transitive closure requirement was
meant as a simple and well understood way to obtain a
\emph{canonical form} for the domain elements, so as to abstract away
from merely syntactic differences.  In~\cite{ShahamKS00}
the specification of all the required abstract semantics operators
is provided, including an operator that is meant to match
the \emph{standard widening} operator defined on the domain of convex
polyhedra~\cite{CousotH78}.
Unfortunately, as pointed out in~\cite{Mine01a,Mine01b}, this operator
is not a widening since it does not provide a convergence guarantee
for the abstract iteration sequence.

The abstract domain of (not necessarily transitively closed) DBMs is
considered in~\cite{Mine01a}.  In this more concrete, syntactic domain
the transitive closure operator behaves as a kernel operator
(monotonic, idempotent and reductive) mapping each DBM into the
smallest DBM (with respect to the component-wise ordering)
encoding the same geometric shape.
As done in~\cite{ShahamKS00}, a widening operator is
also defined in~\cite{Mine01a} and it is observed that this widening
``has some intriguing interactions'' with transitive closure,
therefore identifying the divergence issue faced in~\cite{ShahamKS00}.
This observation has led to the conclusion that
``fixpoint computations \emph{must} be performed''
in the lattice of DBMs, without enforcing transitive closure \cite{Mine01a}.

\section{Difference-Bound Shapes}

While the analysis of the divergence problem is absolutely correct,
the solution identified in~\cite{Mine01a} is sub-optimal
since, as is usually the case, resorting to a syntactic domain
(such as the one of DBMs) has a number of negative consequences.
To identify a simpler, more natural solution, we first have to
acknowledge that an element of this abstract domain should be a
geometric shape, rather than (any) one of its matrix
representations. To stress this concept, such an element will be
called a \emph{Difference-Bound Shape} (DBS).  A DBS corresponds to
the equivalence class of all the DBMs representing it.  The
implementation of the abstract domain can freely choose between these
possible representations, switching at will from one to the other, as
long as the semantic operators are implemented as expected.
The other step towards the solution of the divergence problem is the
simple observation that a DBS is a convex polyhedron and the set of
all DBSs is closed under the application of the standard widening on
polyhedra. Thus, no divergence problem can be incurred when applying
the standard widening to an increasing sequence of DBSs.

On the other hand, the domain of DBSs is isomorphic to the domain of
transitively closed DBMs considered in~\cite{ShahamKS00}, which
suffers from an actual divergence problem.  A closer inspection
reveals that these two observations are not in contradiction, because
the widening operator defined in~\cite{ShahamKS00} is not
equivalent to the standard widening for convex polyhedra.
In fact, a key requirement in the specification of the standard
widening is that the first argument is described by a non-redundant
system of
\ifthenelse{\boolean{short}}
{%
constraints.
}{%
constraints.\footnote{This requirement was sometimes neglected in recent
papers describing the standard widening; it was recently recalled and
exemplified in~\cite{BagnaraHRZ03,BagnaraHRZ05SCP}.}
}
Thus, using transitively closed DBMs does not work because they
typically contain redundant constraints.
What is needed for a correct implementation of the standard widening
is a minimization procedure mapping a DBM representation into (any)
one of the maximal elements in the corresponding equivalence class:
such a procedure was defined in~\cite{LarsenLPY97} and called
\emph{transitive reduction}.

In summary, the solution to the divergence problem for DBSs is to
apply the standard widening of~\cite{CousotH78} to a transitively reduced DBM
representation of the first argument. It is worth stressing that, from
the point of view of the user, this is a transparent implementation
detail: on the domain of DBSs, transitive reduction is the identity
function, as was the case for transitive closure.

\subsection{On the Precision of the Standard Widening}

The standard widening on DBSs
could result, if used without any precaution, in poorer precision
with respect to its counterpart defined on the syntactic domain of DBMs.
The specification of~\cite{Mine01a} prescribes, for maximum precision,
two constraints on the abstract iteration sequence:
the first one restricts the application of the standard widening to a
transitively closed representation for the second argument (note that,
in this case, no divergence problem can arise);
the second one demands that the first DBM of the iteration sequence
$M_0$, $M_1$, \dots,~$M_i$, \dots is transitively closed.
The effects of both improvements can be obtained also with the
semantic domain of DBSs. As for the first one, this can be applied
as is (since transitive closure is just an implementation detail).
The other improvement can be achieved by applying the
well-known `widening up to' technique defined
in~\cite{HalbwachsPR94,HalbwachsPR97}
or its variation called `staged widening with thresholds'
\cite{BlanchetCCFMMMR02,BlanchetCCFMMMR03,Mine04}:
in practice, it is sufficient to add to the set of `up to' thresholds
all the constraints of $M_0$ that are redundant for the representation
of the corresponding DBS (i.e., those constraints that are removed by
the transitive reduction algorithm).

Further precision improvements can be obtained by applying any delay
strategy and/or the framework defined
in~\cite{BagnaraHRZ03,BagnaraHRZ05SCP}.  In particular, by providing
the widening on DBSs with a finite convergence certificate, it is
possible to lift it to a corresponding widening on the \emph{finite powerset}
of DBSs~\cite{BagnaraHZ03b}. It should be stressed that, in this case,
using the syntactic domain of DBMs may have drawbacks: since different
DBMs may represent the same DBS, the presence of these ``duplicates''
in a finite powerset element may have a negative effect on both
efficiency and precision (e.g., when considering a \emph{cardinality-based}
widening operator).  Also note that, in general, the systematic
removal of these duplicates would interfere with widenings, possibly
compromising the convergence guarantee.

\section{Octagonal Shapes and Beyond}

The abstract domain of DBMs has been generalized in~\cite{Mine01b} so
as to allow for the manipulation of constraints of the form
$a x + b y \leq c$, where $a,b \in \{-1,0,+1\}$, leading to the
definition of the \emph{octagon} abstract
\ifthenelse{\boolean{short}}
{%
domain.
}{%
(octagons were called \emph{simple sections} in~\cite{BalasundaramK89}).
}
Each octagon is represented by using a \emph{coherent} DBM
and the transitive closure algorithm is specialized into a
\emph{strong closure} procedure. All the previous reasoning can be
repeated, leading to the definition of the semantic abstract domain of
\emph{octagonal shapes} together with a correct implementation of the
standard widening. In this case, the transitive reduction
algorithm defined in~\cite{LarsenLPY97} does not eliminate all
redundancies: we will describe a new minimization procedure that takes
into account all the constraint inferences performed by the strong
closure algorithm.

Other examples of weakly-relational numeric domains include the `two
variables per inequality' abstract domain~\cite{SimonKH02}, the
octahedron abstract domain~\cite{ClarisoC04}, and the abstract domain
of template constraint matrices~\cite{SankaranarayananSM05}, as well
as the abstract domain of bounded quotients~\cite{Bagnara97th} and the
zone congruence abstract domain \cite{Mine02}.  As long as their
implementation is based on (extensions of) the transitive closure
algorithm, it is possible to define the corresponding syntactic and
semantic versions. The choice between the two versions mainly depends
on the availability of a reasonably efficient minimization procedure:
in our opinion, all the rest being equal, the semantic versions should
be preferred for their greater elegance and the more natural
integration with domain constructions such as the finite powerset
operator.

\ifthenelse{\boolean{short}}
{%
\newcommand{\noopsort}[1]{}\hyphenation{ Ba-gna-ra Bie-li-ko-va Bruy-noo-ghe
  Common-Loops DeMich-iel Dober-kat Di-par-ti-men-to Er-vier Fa-la-schi
  Fell-eisen Gam-ma Gem-Stone Glan-ville Gold-in Goos-sens Graph-Trace
  Grim-shaw Her-men-e-gil-do Hoeks-ma Hor-o-witz Kam-i-ko Kenn-e-dy Kess-ler
  Lisp-edit Lu-ba-chev-sky Ma-te-ma-ti-ca Nich-o-las Obern-dorf Ohsen-doth
  Par-log Para-sight Pega-Sys Pren-tice Pu-ru-sho-tha-man Ra-guid-eau Rich-ard
  Roe-ver Ros-en-krantz Ru-dolph SIG-OA SIG-PLAN SIG-SOFT SMALL-TALK Schee-vel
  Schlotz-hauer Schwartz-bach Sieg-fried Small-talk Spring-er Stroh-meier
  Thing-Lab Zhong-xiu Zac-ca-gni-ni Zaf-fa-nel-la Zo-lo }

}{%
\newcommand{\noopsort}[1]{}\hyphenation{ Ba-gna-ra Bie-li-ko-va Bruy-noo-ghe
  Common-Loops DeMich-iel Dober-kat Di-par-ti-men-to Er-vier Fa-la-schi
  Fell-eisen Gam-ma Gem-Stone Glan-ville Gold-in Goos-sens Graph-Trace
  Grim-shaw Her-men-e-gil-do Hoeks-ma Hor-o-witz Kam-i-ko Kenn-e-dy Kess-ler
  Lisp-edit Lu-ba-chev-sky Ma-te-ma-ti-ca Nich-o-las Obern-dorf Ohsen-doth
  Par-log Para-sight Pega-Sys Pren-tice Pu-ru-sho-tha-man Ra-guid-eau Rich-ard
  Roe-ver Ros-en-krantz Ru-dolph SIG-OA SIG-PLAN SIG-SOFT SMALL-TALK Schee-vel
  Schlotz-hauer Schwartz-bach Sieg-fried Small-talk Spring-er Stroh-meier
  Thing-Lab Zhong-xiu Zac-ca-gni-ni Zaf-fa-nel-la Zo-lo }

}

\end{document}